\documentclass[12pt]{iopart}

\usepackage{graphicx}
\usepackage{iopams}  
\usepackage{color}

\begin{document}

\title{Multiscale interaction between a large scale magnetic island and small scale turbulence}

\author{M J Choi$^1$, J Kim$^1$, J-M Kwon$^1$, H K Park$^{1,2}$, Y In$^1$, W Lee$^1$, K D Lee$^1$, G S Yun$^3$, J Lee$^2$, M Kim$^2$, W-H Ko$^1$, J H Lee$^1$, Y S Park$^4$, Y-S Na$^5$, N C Luhmann Jr$^6$, B H Park$^1$}

\address{$^1$ National Fusion Research Institute, Daejeon 34133, Korea}
\address{$^2$ Ulsan National Institute of Science and Technology, Ulsan 689-798, Korea}
\address{$^3$ Pohang University of Science and Technology, Pohang, Gyungbuk 790-784, Korea}
\address{$^4$ Columbia University, New York, NY 10027, USA}
\address{$^5$ Seoul National University, Seoul 08826, Korea}
\address{$^6$ University of California at Davis, Davis, CA 95616, USA}

\ead{mjchoi@nfri.re.kr}

\date{\today}

\begin{abstract}
Multiscale interaction between the magnetic island and turbulence has been demonstrated through simultaneous two-dimensional measurements of turbulence and temperature and flow profiles. The magnetic island and turbulence mutually interact via the coupling between the electron temperature ($T_e$) gradient, the $T_e$ turbulence, and the poloidal flow. The $T_e$ gradient altered by the magnetic island is peaked outside and flattened inside the island. The $T_e$ turbulence can appear in the increased $T_e$ gradient regions. The combined effects of the $T_e$ gradient and the the poloidal flow shear determine two-dimensional distribution of the $T_e$ turbulence. When the reversed poloidal flow forms, it can maintain the steepest $T_e$ gradient and the magnetic island acts more like a electron heat transport barrier. Interestingly, when the $T_e$ gradient, the $T_e$ turbulence, and the flow shear increase beyond critical levels, the magnetic island turns into a fast electron heat transport channel, which directly leads to the minor disruption. 
\end{abstract}

\vspace{2pc}

\maketitle

\section{Introduction}

A large scale magnetic island in tokamak plasmas was known to degrade the plasma confinement by increasing the radial transport along the reconnected field line. Recent studies, however, have found that the transport physics near the magnetic island can be much more complicated due to various multiscale interactions between the island and small scale turbulence~\cite{Thyagaraja:2005cy,Diamond:2006hx,Nakajima:2007dn,Militello:2008kb,Wang:2009by,Waelbroeck:2009hz,Muraglia:2009dk}. For example, the helical topology of the magnetic island results in 3D perturbation of the magnetic flux surfaces, and profiles of plasma temperature and density are also modified accordingly. Change of pressure profile will be accompanied by changes of flow profile and turbulent fluctuation~\cite{Connor:2009je,Poli:2009ce}. On the other hand, small scale turbulence driven by the background profile can trigger the onset of a magnetic island through a nonlinear beating~\cite{Ishizawa:2010bh,Muraglia:2011ck,Hornsby:2015da} or affect the nonlinear growth of the island~\cite{Hornsby:2011ib,Muraglia:2017fz}. Multiscale interaction between the magnetic island and turbulence is multi-directional and the transport physics near the magnetic island is complicated. 

This paper focuses on effects of the magnetic island on profiles and turbulence and its consequence for the electron heat transport or the nonlinear stability of the magnetic island. Inside the magnetic island, a pressure profile flattens when the island size grows sufficiently large so that the parallel transport along the reconnected field line becomes dominant over the perpendicular transport~\cite{Fitzpatrick:1995ud}. Outside the magnetic island, a pressure profile can be radially steepened because the magnetic flux surfaces are perturbed to be close to each other~\cite{deVriesPC:1997gz}. Reduction of turbulent fluctuation by loss of the pressure gradient inside the flat magnetic island has been observed in~\cite{Bardoczi:2016gj, Bardoczi:2017im}. Experimental measurements of the poloidal flow in the vicinity of the magnetic island have been reported in~\cite{Ida:2002ga,Ida:2004ix,Zhao:2015gra,Rea:2015he,Estrada:2016gz}, and they found that the flow shear across the island can be important in multiscale interactions and consequently in the transport across the island. Recent simulation studies have predicted multiscale interactions via both pressure and flow profiles. They made detail observations such as the localized turbulence distribution~\cite{Connor:2009je,Poli:2009ce,Hornsby:2010fh,Poli:2010hy,Navarro:2017ei,Izacard:2016de} or the poloidal vortex flow around the magnetic island~\cite{Ishizawa:2009es,Hornsby:2010fh,Poli:2010hy,Navarro:2017ei,Hu:2016kea}. The turbulence level is expected to be insignificant across the O-point region probably due to small pressure gradient inside the magnetic island and the strong flow shear outside the magnetic island~\cite{Poli:2009ce,Hornsby:2010fh,Poli:2010hy,Navarro:2017ei}. The turbulent transport is only significant close to the X-point~\cite{Poli:2009ce,Hornsby:2010fh,Poli:2010hy,Navarro:2017ei,Izacard:2016de}. Simultaneous experimental measurements of turbulence and flow in a two-dimensional (2D) space are required to fully validate those multiscale interactions in numerical simulations.

In this paper, the $T_e$ profile, the $T_e$ turbulence, and the poloidal flow near the $m/n=2/1$ magnetic island ($m$ and $n$ are the poloidal and toroidal mode number, respectively) are measured directly and simultaneously in 2D space for the first time. Both the $T_e$ and flow profiles altered by the magnetic island are indeed important in multiscale interactions. The two-dimensional $T_e$ turbulence distribution is determined by the combined effect of the $T_e$ gradient (turbulence drive) and the poloidal flow (turbulence suppression/convection). In particular, when the reversed poloidal flow forms around the magnetic island, the steepest $T_e$ gradient is obtained in the inner region ($r<r_{si}$ where $r_{si}$ represents the inner separatrix of the magnetic island) and the magnetic island acts more like a barrier of the electron heat transport until the transport bifurcation occurs. In section~\ref{s2}, multiscale interaction in the reversed poloidal flow state is described and compared qualitatively with previous studies. In section~\ref{s3}, coupled evolution of the $T_e$ gradient, the $T_e$ turbulence, and the poloidal flow towards the reversed flow state and the transport bifurcation phenomena are discussed. Summary and conclusion are given in section~\ref{s4}.

\begin{figure}[t]
\centering
\includegraphics[keepaspectratio,width=0.6\textwidth]{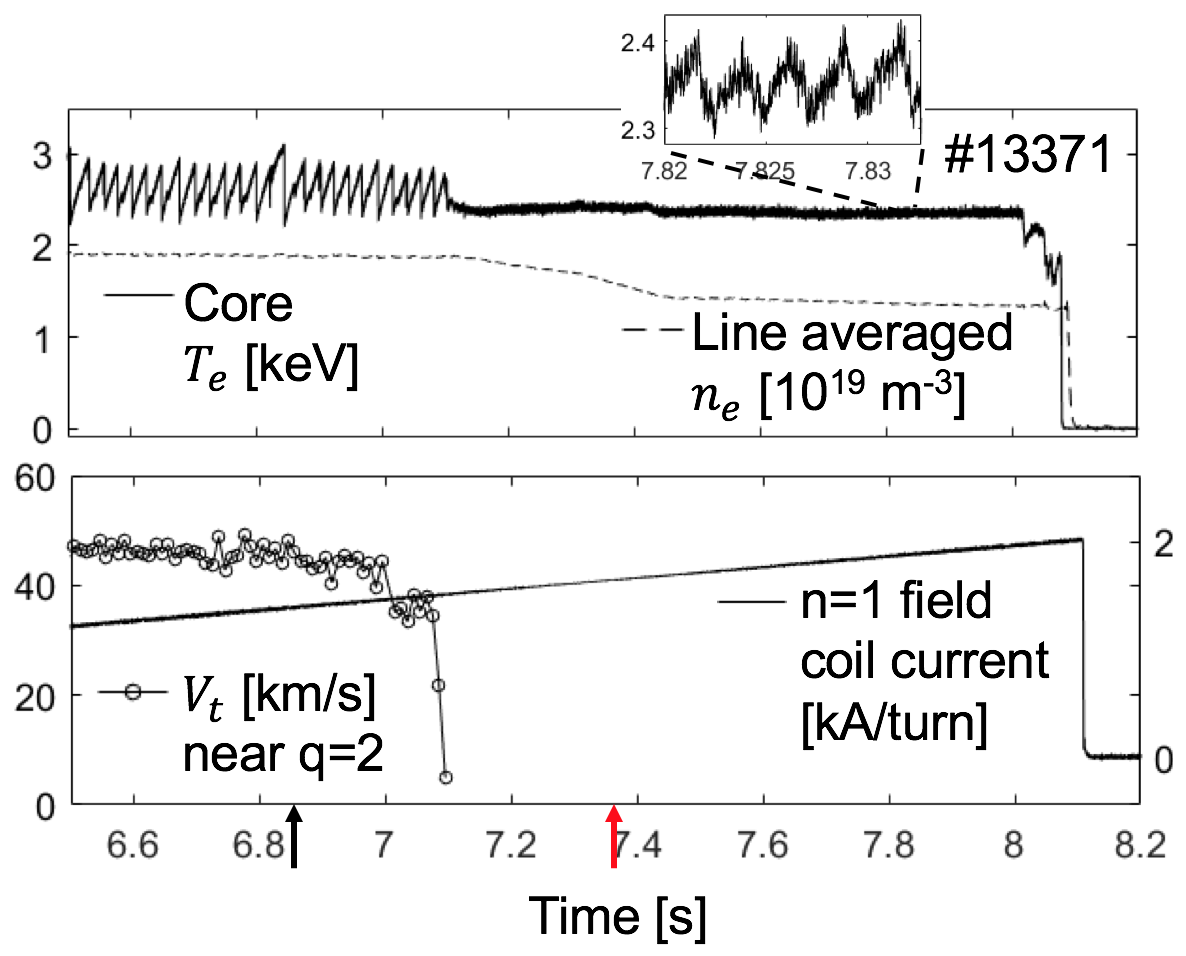}
\caption{(color online). Time traces of the electron temperature ($T_e$), the line averaged density ($n_e$), the toroidal flow speed ($V_t$), and the $n=1$ field coil current per turn in the plasma \#13371. Black and red arrows indicate two time points of w/o and w/ the magnetic island, respectively.}\label{figOV}
\end{figure}

\section{Multiscale interaction in the reversed poloidal flow state~\label{s2}}
\subsection{Experimental set-up}

In the Korea Superconducting Tokamak Advanced Research (KSTAR; major radius $R = 180$~cm and minor radius $a =$~50 cm) experiment \#13371, the plasma was heated by 1 MW neutral beam injection and kept in the low confinement mode with the plasma current $I_p = 0.7$ MA, the safety factor at the 95\% magnetic flux surface $q_{95}\sim 4.6$, and the Spitzer resistivity $\eta \sim 1.4\times 10^{-7}$. The non-rotating $m/n=2/1$ magnetic island was induced by an external $n=1$ resonant magnetic perturbation (RMP) field. Coil current for the $n=1$ RMP field was increased in time as shown in Fig.~\ref{figOV}, and above a critical threshold value the $n=1$ field penetrates deep into the plasma. The toroidal flow speed ($V_t$) near the $q = 2$ region measured by the charged exchange spectroscopy (CES)~\cite{Lee:2011cb} dropped to almost zero within the measurement error ($\pm 5$ km/s) during the penetration. The core electron temperature from the electron cyclotron emission (ECE) diagnostics indicates that the sawtooth crash became very frequent and small~\cite{Piron:2016fa}. A slow decrease in the line averaged electron density, often referred to as the density pump-out, was also observed. The major disruption occurs with the continuously increased $n=1$ field~\cite{Hender:1992wq}.

For measurements of the $T_e$ profile, the $T_e$ turbulence, and the poloidal flow, the 1D ECE diagnostics and the 2D ECE imaging (ECEI) diagnostics~\cite{Yun:2014kv} were utilized. The ECEI diagnostics was cross-calibrated~\cite{Choi:2016ga} using the axis-symmetric $T_e$ profile from the absolutely calibrated ECE diagnostics and the EFIT reconstructed equilibrium~\cite{Park:2011go} in the period w/o the magnetic island in Fig.~\ref{figOV}. The poloidal flow velocity could be deduced from the vertical pattern velocity ($v_{pt}$) ~\cite{Lee:2016bl, Lee:2016kua} estimated using two vertically adjacent ECEI channels. A spatial resolution of the ECEI diagnostics is close to 2 cm in both radial and vertical directions and a temporal resolution is 2~$\mu$s. Note that effects of the relativistic shift, the Doppler broadening, and finite poloidal field~\cite{Rathgeber:2013ej} for the radial channel positions are more or less canceled out in this plasma condition, and the cold resonance positions could be used. In the outer region ($r > r_{so}$ where $r_{so}$ means the outer separatrix of the magnetic island), $T_e$ measurement is uncertain because the ECE diagnostic capability becomes marginal. In terms of the optical depth ($\tau$)~\cite{Hutchinson:2002ws}, it is close to or less than 1 in the outer region while close to 3 in the inner region and in between 1 and 3 inside the magnetic island. 

\begin{figure}[t]
\centering
\includegraphics[keepaspectratio,width=0.6\textwidth]{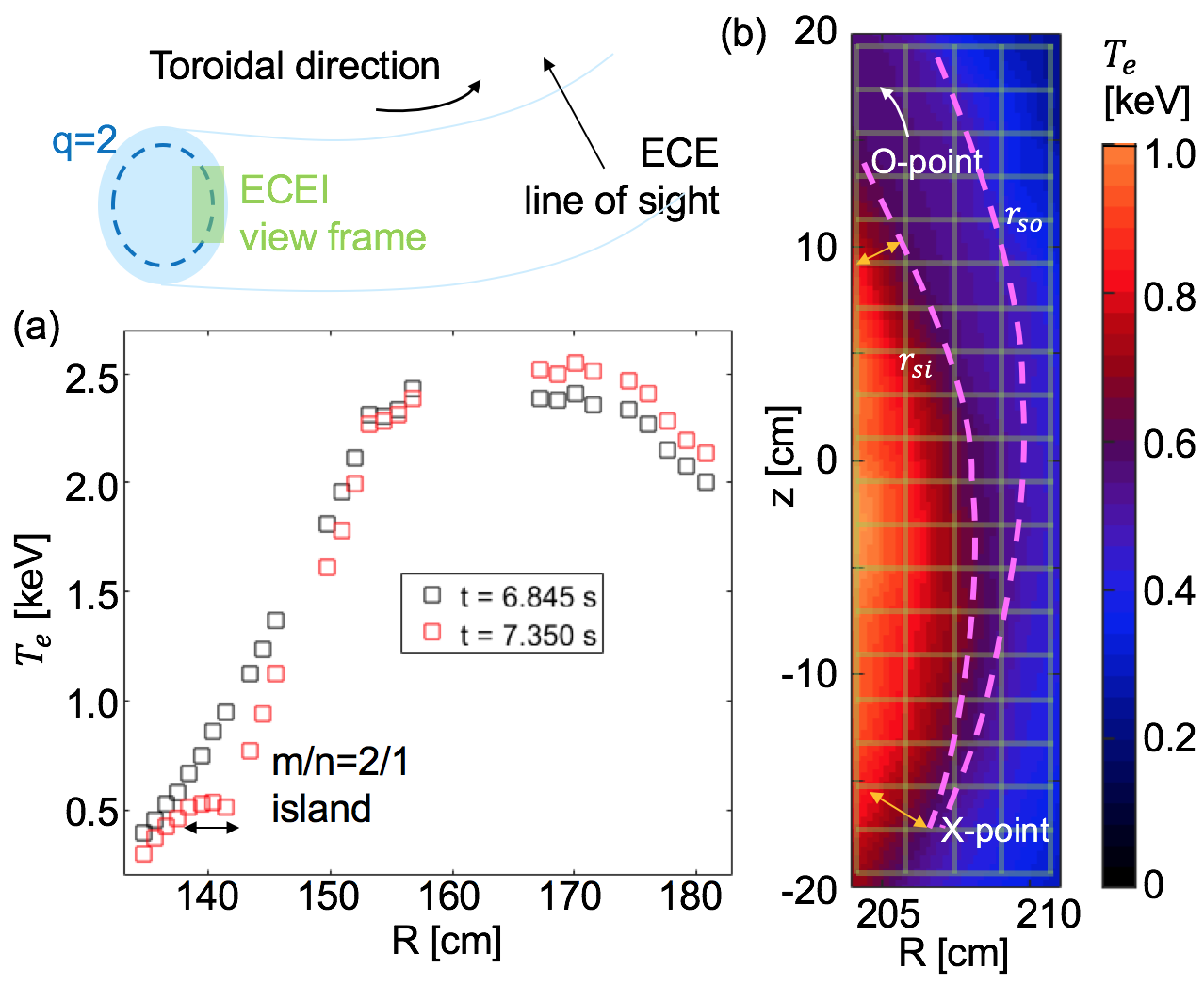}
\caption{(color online). The ECE and ECEI diagnostics are installed at different toroidal ports separated by 90 degree. (a) Radial time averaged $T_e$ profiles w/o (black) and w/ (red) the $m/n=2/1$ magnetic island in the plasma \#13371. (b) The 2D cross-calibrated $T_e$ profile with the estimated separatrix of the magnetic island (purple dashed line).}\label{figPF}
\end{figure}

\subsection{The $T_e$ profile with the magnetic island}

When the $m/n=2/1$ magnetic island is induced, the $T_e$ profile is altered along the magnetic topology of the island and it is no longer axis-symmetric. The radial $T_e$ profiles measured by the ECE diagnostics in the high field side and the 2D $T_e$ profile by the ECEI diagnostics in the low field side at different toroidal angles are shown in Fig.~\ref{figPF}. The $T_e$ profile inside the magnetic island flattens probably due to the fast parallel transport along the reconnected field line~\cite{Fitzpatrick:1995ud} and/or the negligible turbulence spreading~\cite{Navarro:2017ei}. The full width of the magnetic island ($W$) will be close to or larger than 5 cm which is larger than the typical critical width ($W_c \sim 1.0$ cm) for the $T_e$ flattening in the KSTAR L-mode plasmas~\cite{Choi:2014kj}. Note that the separatrix of the magnetic island in the 2D $T_e$ profile can be roughly estimated by the temporal behavior of the electron temperature. A full 2D measured electron temperature profile and a proper modeling with a synthetic diagnostics are needed to estimate the magnetic island full width accurately~\cite{Choi:2014kj, Bardoczi:2016gjba}, especially when the localized (not uniform) and dynamic turbulence exists around the magnetic island which can affect the perpendicular electron heat transport characteristics~\cite{Agullo:2017ig}. 

In contrast to the flattened $T_e$ profile inside the magnetic island, the $T_e$ profile in the inner region ($r < r_{si}$) becomes more steepened with increase of the core $T_e$ level (Fig.~\ref{figPF}(a)). In particular, the $T_e$ gradient increases towards the O-point region as indicated by widths of the orange arrows in the 2D $T_e$ profile in Fig.~\ref{figPF}(b). More closely packed magnetic flux surfaces due to the magnetic island may induce some local $T_e$ profile modifications~\cite{Poli:2009ce,Poli:2010hy,deVriesPC:1997gz}. In order to understand formation of the global peaked $T_e$ profile, the electron heat transport around the magnetic island needs to be studied with measurements of the $T_e$ turbulence and the poloidal flow as follows. 

\begin{figure}[t]
\centering
\includegraphics[height=0.5\textwidth,width=0.55\textwidth]{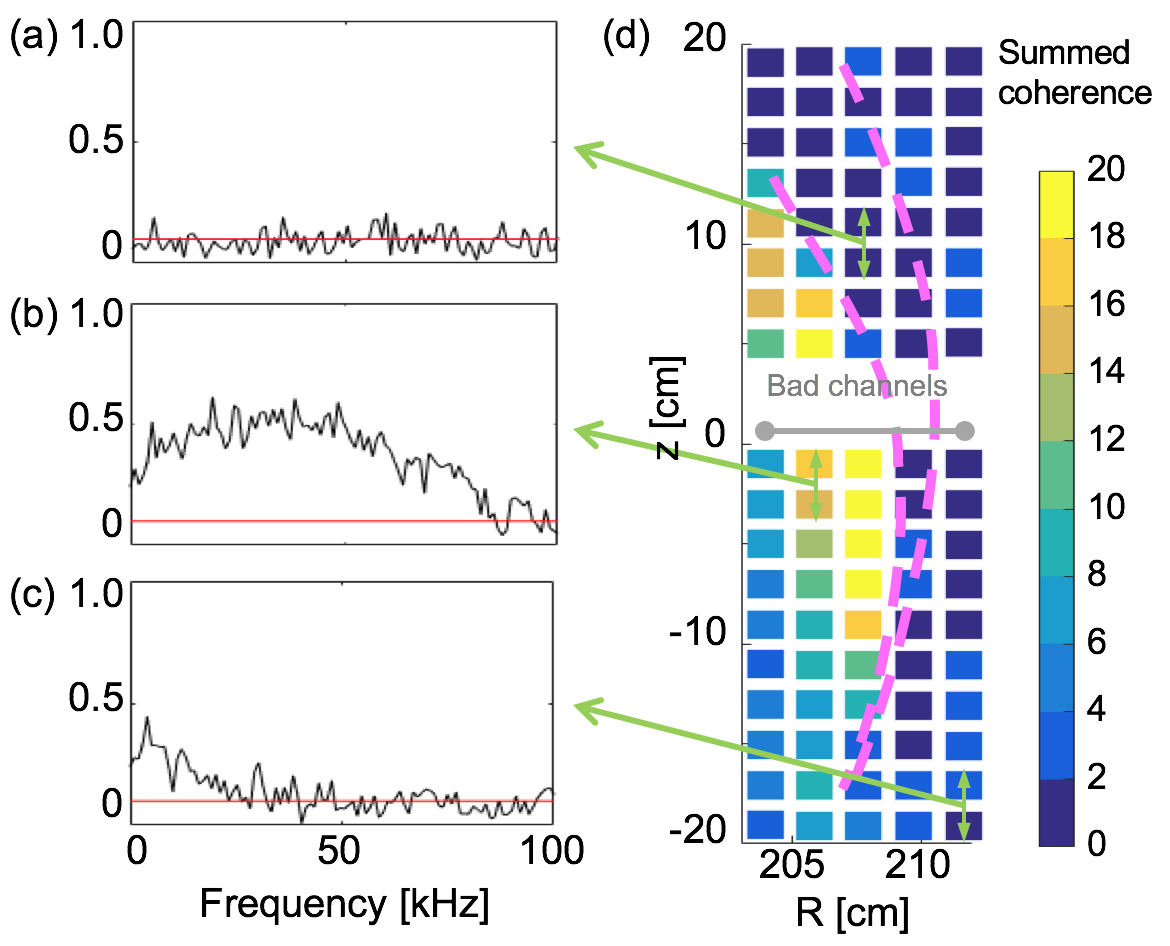}
\caption{(color online). The $\delta T_e / \langle T_e \rangle $ cross coherence (a) inside the magnetic island and in the (b) inner and (c) outer regions in the plasma \#13371. Red line is a significance level. (d) The summed coherence image is obtained using pairs of vertically adjacent ECEI channels. Dashed purple lines indicate the expected magnetic island separatrix. }\label{figTL}
\end{figure}

\subsection{The $T_e$ turbulence and its characteristics}

To estimate the electron turbulent heat transport near the magnetic island, the $T_e$ fluctuations measured by the ECEI diagnostics are analyzed. For example, Figs.~\ref{figTL} (a)---(c) are the cross coherence of $\delta T_e / \langle T_e \rangle \equiv (T_e - \langle T_e \rangle) / \langle T_e \rangle$ where $\langle ~ \rangle$ means the time average. It represent the coherent fraction of the total $\delta T_e / \langle T_e \rangle$ fluctuation power. They are calculated using two vertically adjacent ECEI channels for $t=$~7.35--7.40 s in the plasma \#13371. One inside the magnetic island does not show a significant coherent fluctuation, but the others show some coherent fluctuation power. In the inner region where the $T_e$ gradient is increased significantly, the fluctuation power over a broad frequency band ($0 \le f \le 75$ kHz) is measured clearly. The $T_e$ gradient may be considered as a predominant drive of this turbulent fluctuation. In fact, the coherence increases with the $T_e$ gradient as shown in~Fig.~\ref{figDT}. Note that the weak fluctuation power over a narrow frequency band ($0 \le f \le 30$ kHz) is measured in the outer region. 

\begin{figure}[t]
\centering
\includegraphics[height=0.5\textwidth,width=0.6\textwidth]{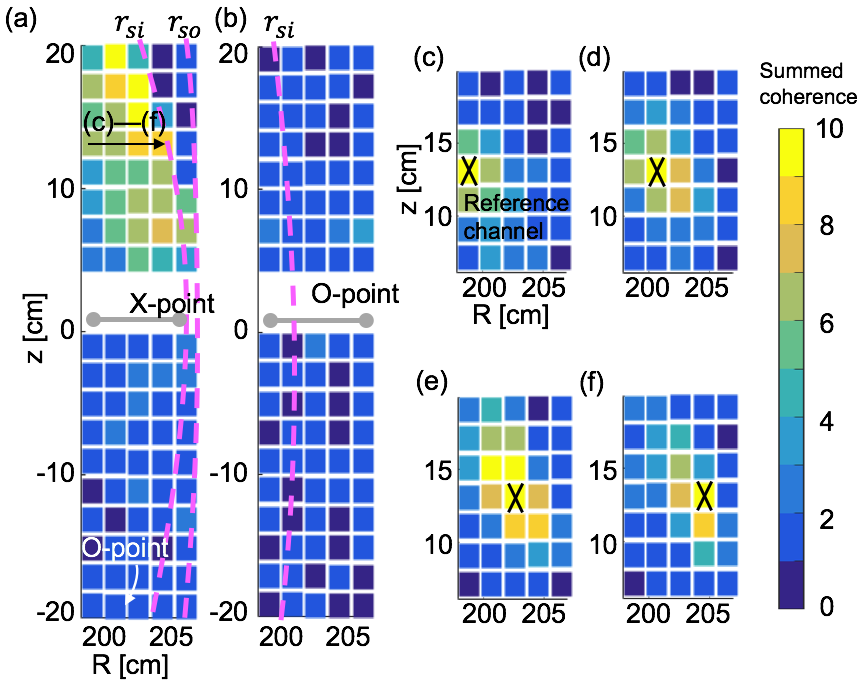}
\caption{(color online). The summed coherence images in the plasma \#15638 for (a) the X-point period and (b) the O-point period are obtained using pairs of vertically adjacent ECEI channels. (c)--(f) The summed coherence images using pairs of a reference channel whose position is indicated by a black cross and the other channels for the X-point period.}\label{fig2D}
\end{figure}

A detail 2D distribution of the $T_e$ turbulence level can be investigated by calculating the summed cross coherence image using more ECEI channels. The cross coherence only above a significance level is summed over a 10--75 kHz band to make the summed coherence image. Note that a 0--10 kHz band was neglected because some channels suffer from 4 kHz electronics noise in this experiment. Each dot in the images in Fig.~\ref{figTL}(d) and Figs.~\ref{fig2D}(a) and \ref{fig2D}(b) represents the summed coherence estimated using the channel at that position and the one below. Note that one row of the ECEI channels had a low signal-to-noise ratio and reliable coherence calculations in two rows near the midplane are not available. The smooth and continuous 2D $T_e$ profile in Fig.~\ref{figPF}(b) is obtained by interpolations. 

The summed coherence image in Fig.~\ref{figTL}(d) shows that the strong $T_e$ turbulence is localized both radially and poloidally in the inner region. It has the maximum close to the inner separatrix of the magnetic island near the X-point. The insignificant ($<2$) summed coherence is observed inside the magnetic island, and weak but meaningful coherence is observed in the outer region. 

The $T_e$ turbulence distribution has been further studied in a similar KSTAR plasma \#15638 in which the toroidal phase of the applied $n=1$ field is slowly varying at the frequency of 2 Hz. In that experiment, both the X-point and O-point regions can be captured in the ECEI view frame in different time periods (20 ms each) and the $\delta T_e / \langle T_e \rangle$ summed coherence images are obtained as shown in Figs.~\ref{fig2D}(a) and \ref{fig2D}(b), respectively. 

The summed coherence is insignificant everywhere for the O-point period, which implies the small turbulent electron heat transport there~\cite{Bardoczi:2016gj,Bardoczi:2017im,Poli:2009ce,Hornsby:2010fh,Poli:2010hy,Izacard:2016de,Navarro:2017ei}. For the X-point period, it is found that the significant coherence is not only localized but also poloidally asymmetric against the X-point. Note that this localized turbulence follows the X-point, which is rotating with the RMP field, with a constant poloidal shift.

The localized asymmetric turbulence near the X-point region strongly suggests that the $T_e$ gradient is not the only control parameter in growth of the $T_e$ turbulence. The poloidal flow can be important as it will be discussed in next section. In fact, the poloidal shift of the turbulence with respect to the X-point coincides with the direction of the local poloidal flow~\cite{Poli:2009ce,Izacard:2016de,Hu:2016kea}. This locality of the island-associated $T_e$ turbulence is consistently observed in other experiment~\cite{Lucas:2016tr}. Although it is beyond the scope of this paper, the radial locality may also imply that the magnetic island itself can be important in driving the turbulence via a direct nonlinear coupling~\cite{Ishizawa:2009es, Wang:2009by, Hu:2014ksa}. 

At this point, it would be helpful to provide some quantitative characteristics of the $T_e$ turbulence such as the rms amplitude, correlation lengths, and the poloidal wavenumber. Firstly, the rms amplitude can be measured by integrating the cross power spectral density between vertically adjacent channels over a 10--75 kHz band, and the maximum turbulence rms amplitude is about $2.5 \pm 0.25 \%$ in the plasma \#13371. Note that the 2D rms amplitude has the almost same distribution with the summed coherence image in Fig.~\ref{figTL}(d). Next, the summed coherence images in Figs.~\ref{fig2D}(c)--(f) are calculated especially for estimation of correlation lengths in the plasma \#15638. Pair of a fixed reference channel (indicated by a black cross) and other channel are used to estimate the correlation length defined as a range of the significant summed cross coherence. The correlation length is found to be not uniform and has a finite poloidal (2--6 cm) and radial (2--3 cm) range. Note that it is a little larger than the radial correlation lengths of density fluctuation across the magnetic island in ~\cite{FernandezMarina:2017el}. Lastly, the poloidal wavenumber of the $T_e$ turbulence can be estimated from the cross phase ($\Delta \Theta$) between vertically adjacent ECEI channels. Figs.~\ref{figVP}(a) and \ref{figVP}(b) represent the vertical ECEI cross phase measured in the inner and outer region of the plasma \#13371, respectively. Fluctuations in a range of $k_\theta \rho_i \approx \frac{\Delta \Theta }{ \Delta z} \rho_i \le 0.4$ were revealed in the most channels in the inner region and in some channels in the outer region where $\rho_i$ is the ion gyroradius. The vertical distance between two adjacent channels ($\Delta z$) was set to be about 2 cm and detectable poloidal wavenumber is roughly limited to $k_\theta \rho_i \le 0.4$ in this experiment. 

\subsection{The poloidal flow}

\begin{figure}[t]
\centering
\includegraphics[height=0.45\textwidth,width=0.55\textwidth]{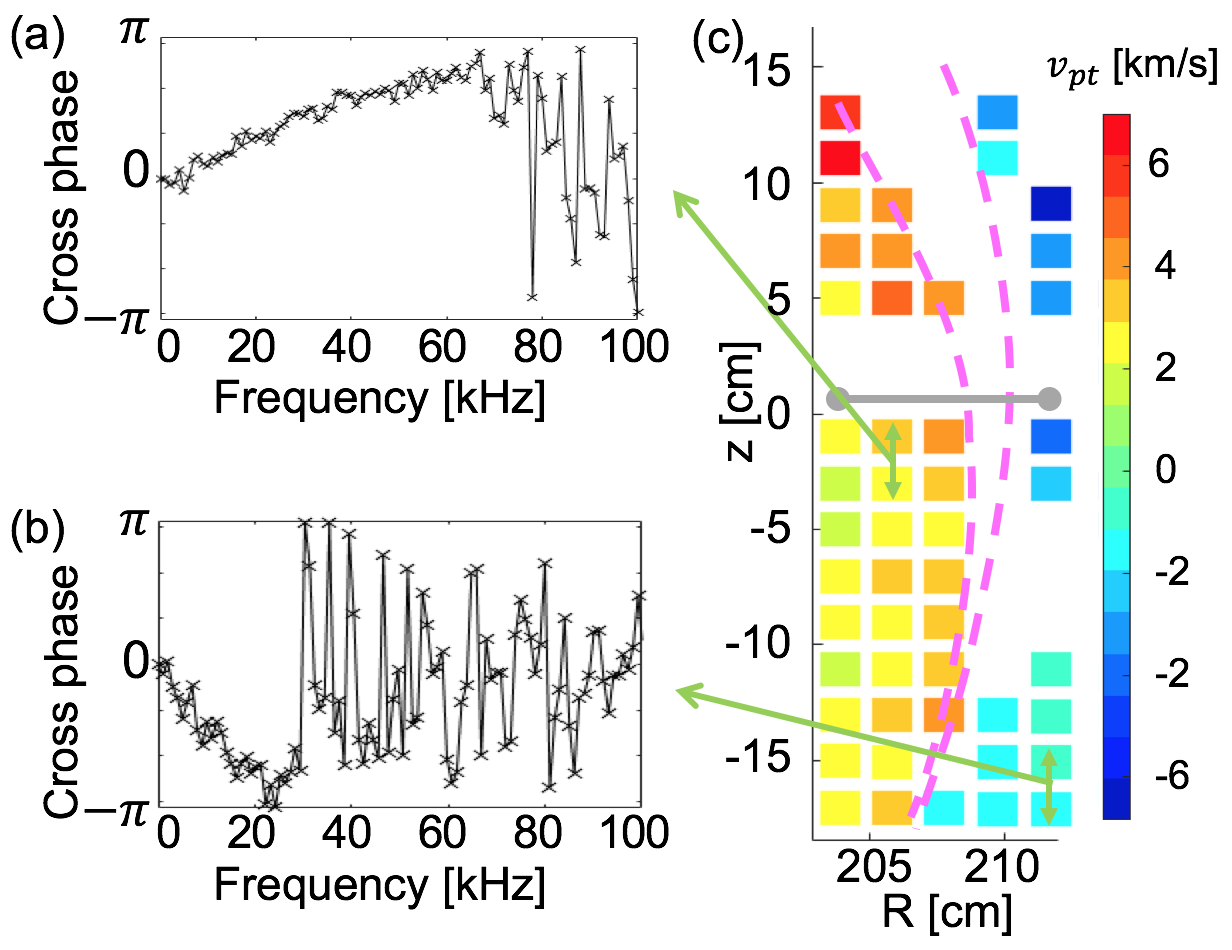}
\caption{(color online). Cross phase between two vertically adjacent ECEI channels measured in the (a) inner and (b) outer regions in the plasma \#13371. (c) The 2D $v_{pt}$ profile is measured using the coherent cross phase.} \label{figVP}
\end{figure}

Using the slope of the coherent vertical ECEI cross phase, the vertical group velocity in the laboratory frame, or the pattern velocity ($v_{pt}$), can be measured~\cite{Lee:2016bl, Lee:2016kua}. Fig.~\ref{figVP}(c) shows 2D measurement of the $v_{pt}$ near the magnetic island for $t=$~7.35--7.40 s in the plasma \#13371. Note that the $v_{pt}$ measured with uncertainty less than $0.8$~km/s is only shown. For the accurate $v_{pt}$ measurement, the ECEI data should have sufficient power of the coherent fluctuation and record length (at least 50 ms). 

In this state, the $v_{pt}$ in the inner region is positive (a counter clockwise or the electron diamagnetic direction), and its speed is radially peaked near the separatrix of the magnetic island. More importantly, it is not uniform in poloidal direction, i.e. it increases toward the O-point region. Therefore, the positive radial shear of the poloidal flow ($dv_{pt}/dr \ge 10^5 s^{-1}$) forms in the inner region and it also increases toward the O-point region. This $v_{pt}$ behavior is consistent with the numerical simulation results~\cite{Poli:2009ce,Hornsby:2010fh,Poli:2010hy,Navarro:2017ei}, and can explain that the $T_e$ turbulence is not detected and the steep $T_e$ profile is maintained near the O-point region. In addition, the $v_{pt}$ is reversed across the magnetic island and the strong negative radial shear of the poloidal flow (-$dv_{pt}/dr > 10^5 s^{-1}$) develops across the island. Although it was not possible to measure the flow inside the flat and quiet magnetic island, the poloidal flow around and inside the island is expected to have a vortex structure~\cite{Ida:2004ix,Estrada:2016gz,Hornsby:2010fh,Poli:2010hy,Navarro:2017ei,Hu:2016kea}. This 2D sheared poloidal flow can prohibit a turbulent eddy from developing across the magnetic island~\cite{Wang:2009jua} and from spreading into the island~\cite{Navarro:2017ei}.

Origin of the poloidal flow is not clearly understood yet. The toroidal flow decreases significantly after the field penetration and its contribution would be negligible in the $v_{pt}$. The diamagnetic drift may serve as a nearly uniform and small background, considering evolution of poloidal flows in Fig.~\ref{figDT}(d). Note that from $t = 7.50$~s to $t = 7.55$~s the plot E showed a drastic change ($+4$~km/s) which is hardly explained by the $10\%$ increase of the electron temperature gradient. The $E \times B$ flow~\cite{Poli:2009ce,Ishizawa:2009es,Hornsby:2010fh,Poli:2010hy,Navarro:2017ei,Hu:2016kea,Ciaccio:2015wb,Leconte:2017jn} or zonal flow driven by the turbulence itself~\cite{Li:2009fh} may play a role in the measured $v_{pt}$.

\section{The poloidal flow reversal and the transport bifurcation ~\label{s3}}

In previous experiments, the applied RMP field strength keeps increasing in time, and it may not be appropriate to study the temporal coupled evolution between the $T_e$ gradient, the $T_e$ turbulence, and the poloidal flow.  In the experiment \#16150, the constant and non-rotating $n=1$ RMP field is applied and the plasma is maintained in the mode-locking state without the major disruption. A repetitive minor disruption is observed as the plasma evolves in time with the constant RMP field, and the coupled evolution is studied for a single minor disruption cycle.

\begin{figure}[t]
\centering
\includegraphics[width=1\textwidth,height=0.45\textwidth]{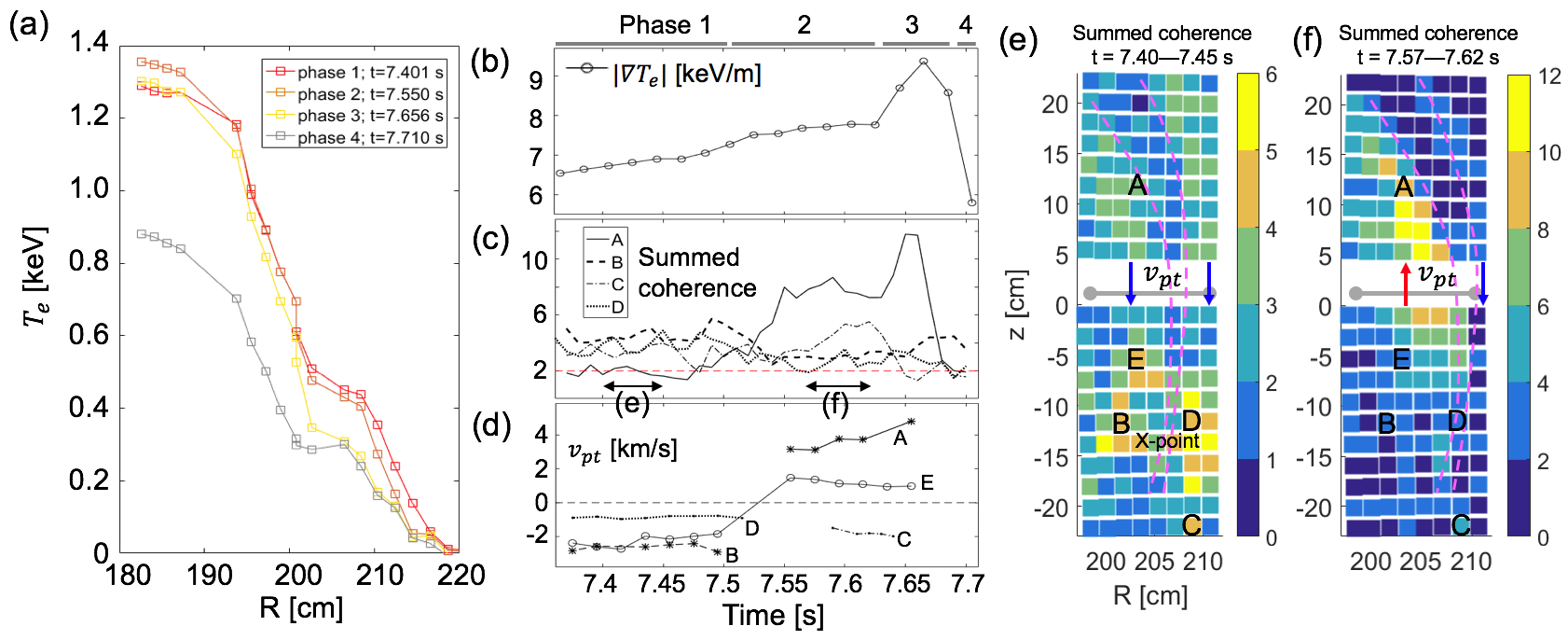}
\caption{(color online). (a) The $T_e$ profile, (b) the $T_e$ gradient in the inner region, (c) the summed $\delta T_e / \langle T_e \rangle$ coherence at different positions, and (d) the $v_{pt}$ at different positions in the plasma \#16150. } \label{figDT}
\end{figure}

Four distinctive phases are observed during a single minor disruption cycle as illustrated in $T_e$ profiles in Fig.~\ref{figDT}(a). The temporal evolutions of the $T_e$ gradient in the inner region, the $T_e$ turbulence level (the summed cross coherence) at different positions (A, B, C, and D), and the poloidal flow ($v_{pt}$) at different positions (A, B, C, D, and E) are shown in Figs.~\ref{figDT}(b)--(d), respectively. Note that the summed coherence in Fig.~\ref{figDT}(c) and the summed coherence image for the phase 1 (Fig.~\ref{figDT}(e)) and the phase 2 (Fig.~\ref{figDT}(f)) are obtained using pair of vertically adjacent ECEI channels and the cross coherence over a 0--60 kHz band in which there is no electronics noise in this experiment.

In the initial phase 1, the $T_e$ gradient in the inner region is not very steep but increasing in time as shown in Fig.~\ref{figDT}(b). The summed cross coherence image in Fig.~\ref{figDT}(e) shows that the coherent fluctuation power is relatively weak but peaked across the X-point of the magnetic island. The negative poloidal flow is sheared across the X-point as shown in the $v_{pt}$ measurement at B, D, and E in Fig.~\ref{figDT}(d), but near the X-point it seems not to be strong enough to affect the turbulence distribution. Although the accurate entire two-dimensional flow measurement was not available in this phase due to the marginal turbulent fluctuation power except the X-point region, the localized turbulence near the X-point implies that the flow shear may be effective beyond the X-point region.

The transition from the phase 1 to the phase 2 involves with a rapid increase of the $T_e$ gradient, i.e. $T_e$ increases at the core region and decreases slightly in the $q\ge2$ region, as well as changes of the 2D $T_e$ turbulence level and poloidal flow. Note that the 2D estimated magnetic island geometry (indicated by the dashed purple line) is also perturbed as seen decrease of the summed coherence at D, which might involve change of the island full width. The line averaged density is nearly constant in the transition and decreases by a few percent later in phase 3. The electron density profile measured by the Thomson scattering system~\cite{Lee:2010el} becomes a little broader but it is not clear due to the unsatisfactory measurement condition. 

In the phase 2, the 2D $T_e$ turbulence distribution is changed as shown in Fig.~\ref{figDT}(f) as Fig.~\ref{figTL}(d), and the reversed poloidal flow forms as shown in the $v_{pt}$ measurement at A, C, and E in Fig.~\ref{figDT}(d) as Fig.~\ref{figVP}(c). The poloidal flow reversal can be originated from change in $v_{E \times B}$ around the magnetic island by the nonlinear resonant low $n$ electrostatic mode~\cite{Ishizawa:2009es,Hornsby:2010fh,Poli:2010hy,Navarro:2017ei} or the response potential to the magnetic perturbation in the initial shear flow~\cite{Hu:2016kea}. The strongly sheared flow developed across the magnetic island can prohibit the turbulence growth or convection across the X-point~\cite{Wang:2009jua,Navarro:2017ei} and shift the $T_e$ turbulence level upwards in the inner region as observed in Fig.~\ref{figDT}(c) and \ref{figDT}(f), which can explain the $T_e$ gradient increase in phase 2.

A sudden decrease of electron temperature in the $q\ge2$ region occurs in the phase 3 through some unknown process (possibly related to edge modes), which leads to a jump in the $T_e$ gradient and the $T_e$ turbulence in the inner region. In addition, the stronger radial shear of the poloidal flows in the inner region (difference between A and E in Fig.~\ref{figDT}(d)) and across the X-point (difference between A and C in Fig.~\ref{figDT}(d)) are observed. 

When all the $T_e$ gradient, the $T_e$ turbulence, and the flow shear increase significantly, a massive fast ($\sim 100~\mu s$) $T_e$ collapse occurs. Note that the $T_e$ profile collapses in two steps, i.e. the local $q\approx2$ region collapse and the $q \le 1$ region collapse, which is very similar with the large minor disruption in~\cite{Choi:2016jz} where the RMP field was not applied. The role of the magnetic island has been changed from a barrier of the electron heat transport (from phase 1 to phase 3) to a fast channel (from phase 3 to phase 4). The observed transport bifurcation may be relevant to either the bifurcation observed in~\cite{Ida:2016ka}, secondary innstabilities~\cite{Li:2014jg, Muraglia:2009dk}, or the vortex flow shear destabilization of the long wavelength fluctuation~\cite{Hu:2016kea}.

\section{Summary and conclusion ~\label{s4}}

The 2D profiles of $T_e$ and poloidal flow and the 2D $T_e$ turbulence distribution are closely coupled around the magnetic island. The magnetic island and turbulence mutually interact via this coupling which has a critical effect on the electron heat transport. The magnetic island can play as either a barrier or a fast channel of the electron heat transport. 

In particular, the magnetic island acts more like an electron heat transport barrier when the poloidal flow is perturbed to have a strongly sheared profile. The speed of the flow is peaked near the separatrix of the magnetic island increasing towards the O-point region. The positive flow shear in the inner region would suppress the $T_e$ turbulence around the O-point region, and the $T_e$ turbulence level is only significant in the narrow region close to the X-point region. The negative flow shear across the magnetic island would prevent a turbulent eddy from growing across the X-point and from spreading into the island. In this state, the poloidal flow developed around the magnetic island seems to regulate the electron turbulent heat transport across the magnetic island. 

However, when the $T_e$ gradient, the $T_e$ turbulence, and the flow shear exceed critical levels, the transport bifurcation occurs and a massive heat transport event follows. The role of the magnetic island on the electron thermal transport is more complicated than a direct thermal loss channel. 

This experiment clearly demonstrates multiscale nonlinear interaction between a large scale magnetohydrodynamic instability and small scale turbulence and its importance on the electron thermal transport. It may provide some physical insights to understand the internal transport barrier formation~\cite{Waelbroeck:2001dl,Wolf:2003wx,Chung:ug} or the RMP edge localized mode suppression~\cite{Evans:2006hra,Jeon:2012hu}. More researches focused on the validation of the ongoing gyrokinetic simulation with the experimental observation will be done in near future. 

\section*{Acknowledgement}
One of the authors (M. J. C.) acknowledges helpful discussions with Dr. J. Seol, Dr. J.-H. Kim, Dr. M. Leconte, Dr. S. Zoletnik, and Dr. L. Bard{\'o}czi. This work is supported by Korea Ministry of Science, ICT and Future Planning under KSTAR project (Contract No. OR1509) and under NFRI R\&D programs (NFRI-EN1741-3), and also partly supported by NRF Korea under Grant No. NRF-2014M1A7A1A03029865 and NRF-2014M1A7A1A03029881. 

\section*{References}

\bibliographystyle{iopart-num}

\providecommand{\newblock}{}

\end{document}